\documentclass{elsart5p}
\usepackage{amssymb}
\usepackage[dvips]{graphicx}
 \journal{Surface Science}
\date{}

\begin{document}
\begin{frontmatter}
\title{Polarization anisotropy in the optical properties of silicon ellipsoids}
\author{F. Trani}
\address{Dipartimento di Scienze Fisiche, Universit\`a di Napoli ``Federico II,'' Complesso Universitario Monte S. Angelo, Via Cintia, I-80126 Napoli, Italy}

\begin{abstract}
A new real space quantum mechanical approach with local field effects included is applied to the calculation of the optical properties of silicon nanocrystals. Silicon ellipsoids are studied and the role of surface polarization is discussed in details. In particular, surface polarization is shown to be responsible for a strong optical anisotropy in silicon ellipsoids, much more pronounced with respect to the case in which only quantum confinement effects are considered. The static dielectric constant and the absorption spectra are calculated,
showing that the perpendicular and parallel components have a very different dependence on the ellipsoid aspect ratio. Then, a comparison with the classical dielectric model is performed, showing that the model only works for large and regular structures, but it fails for thin elongated ellipsoids.
\end{abstract}

\end{frontmatter}

\section{Introduction}  \label{sec:intro}
Elongated silicon nanocrystals (Si-nc) are expected to play an important role in the current process of miniaturization of technology\cite{bhushan04}. Small silicon quantum wires are already fabricated in labs\cite{ma03} and their use in gas sensing activities\cite{cui01a} and nano-electronics\cite{she06} opens new scenarios for the next future. Also, elongated silicon nanocrystallities are the building blocks responsible for both the photoluminescence\cite{kovalev96} (PL) and the recently observed polarized optical gain\cite{cazzanelli04} from Porous Silicon (PS) samples.
Elongated Si-nc are very interesting objects for the fundamental physics, too. Indeed, the role of the surface polarization on the electronic properties of Si-nc has been recently raising a wide interest\cite{ninno06,delerue03}.
Surface polarization effects (SPEs) are extremely important for Si-nc optical properties. In particular, SPEs are responsible for a strong optical anisotropy in elongated structures, as it has been shown for carbon nanotubes\cite{li01,marinopoulos03}, Si ellipsoids and wires\cite{kovalev96, bruneval05} and Si clusters\cite{idrobo06}.
From the theoretical point of view, it is computationally heavy to include SPEs in an atomistic quantum mechanical framework. Only recently first-principles calculations which include local field effects (LFEs) have been performed on small Si-nc and very thin Si quantum wires\cite{bruneval05,gatti05}. 

In this work we use an empirical tight binding scheme (TB) to study the role of LFEs in Si-nc. Indeed, as we have recently shown, LFEs in Si-nc are mostly due to SPEs\cite{ninno06}, which are therefore taken into account in the present treatment. The paper is organized in the following way. In section \ref{sec:classical} the classical dielectric model is briefly illustrated, in section \ref{sec:method} the TB method is drawn, in section \ref{sec:results} we show the results obtained for ellipsoidal Si-nc, both the static dielectric constant and the absorption spectra, upon changing the geometrical anisotropy of the structures.

\section{The classical model} \label{sec:classical}
In a semiconductor structure embedded into a background with a different dielectric constant, an external electric field causes a charge accumulation on the surface. The field inside the structure is the sum of the external field, and a depolarization field due to this charge accumulation, the latter being at the origin of SPEs. SPEs are very important in situations in which there is a high dielectric mismatch across the surface and the depolarization field is huge. From a classical point of view, SPEs are usually taken into account by solving the Maxwell equations with suitable boundary conditions\cite{landau}. The basic assumption is that both the structure and the background behave as ideal dielectric media, characterized by a position independent dielectric constant. This gives the classical dielectric model. The starting point is the dielectric constant of the structure, that is either the bulk one, when quantum confinement effects (QCEs) are negligible, or a size dependent Si-nc dielectric constant, derived, for example, from a Penn-like model or a semiempirical approach\cite{trani05}. In the following we start from the independent particle dielectric constant derived from our tight binding method. We refer to this model as \textit{semiclassical}, since QCEs are correctly taken into account, but SPEs are included using classical electrostatics. It is worth stressing that in this model SPEs enter the polarizability of a structure only through the geometrical shape and the dielectric mismatch across the surface, but they are independent of the nanocrystal size. 

The dielectric model has been shown working well for huge structures with a quite regular shape. For structures small and with a complex shape, the model is expected to fail and more accurate methods have to be used. In the next section we are going to illustrate an atomistic quantum mechanical approach for the inclusion of SPEs into the Si-nc optical properties. From a comparison between both the approaches, we can give an indication about the range of validity of the semiclassical model.

\section{The method} \label{sec:method}
The empirical tight binding method (TB) is a very useful tool in the study of structures with thousand of atoms. It is based on the expansion of the nanocrystal wavefunctions into a localized basis set and is computationally very light.
For the study of Si-nc we use an $sp^3$ orthogonal parametrization, with interactions up to third nearest neighbors, which has been shown to give a nice agreement with the experimental energy gaps and absorption data and well compares to more sophisticated first-principles calculations. All the details of the method can be found in Refs.\cite{trani05,trani04}, and reference therein.

In the literature, the Si-nc optical properties are usually studied within the independent particle random phase approximation (RPA), and only recently beyond-RPA calculations have been presented\cite{benedict03}, but they mostly concern very small clusters.
We use a TB formulation of the linear response theory with LFEs included (RPA+LF), which is derived from the extension of the Hanke and Sham formalism to confined structures\cite{hanke74,hanke79}. The method is based on a real space approach that is very efficient for the study of confined structures. Moreover, the approach is very simple and has been well-tested for Si-nc\cite{delerue03,delerue97}.

The macroscopic dielectric constant of a bulk system is defined starting from the calculation of the reciprocal space inverse dielectric matrix\cite{hanke74}. LFEs are related to the off-diagonal components of the reciprocal space dielectric matrix. Within the independent particle RPA, the off-diagonal components, and therefore LFEs, are neglected. By including the off-diagonal components into the calculation, we retrieve the RPA+LF approximation. As it is known, in bulk Si LFEs give a negligible contribution to the absorption\cite{hanke79}. The situation is very different in Si-nc, since SPEs (which are a particular kind of LFEs), play a strong role, as we have recently shown in the case of point charge screening\cite{ninno06}.

For the Si-nc optical properties calculations, LFEs are included into a tight binding scheme using a real space approach. The starting point is the diagonalization of the Si-nc tight binding hamiltonian. Using an expansion into the atomic orbital basis set, the linear response theory can be developed according to the method proposed in Ref.\cite{hanke74} for the bulk, with the basic important difference that the Si-nc wavefunctions replace the bulk Si Bloch functions. In such a way, a fully real space approach can be performed, and there is no need to introduce a fictitious supercell. Using the diagonal approximation for the position matrix elements (see Ref.\cite{trani05} and references therein), the following expression can be obtained for the frequency dependent dielectric tensor\cite{hanke74}
\begin{equation}
\label{eq:tb}
\varepsilon _{\beta,\gamma} \left(\omega\right) = 1 - \frac{4\pi e^2}{\Omega} \sum _{i,j }\mathbf{R}_i^{\beta} S\left(\omega\right)_{ij} \mathbf{R}_j^{\gamma}.
\end{equation}
Here, $\Omega$ is the Si-nc volume, $\mathbf{R}_i^{\beta}$'s are the positions of atoms composing the structure, $\beta$ and $\gamma$ are cartesian components, $S_{ij}$ is the real space screened polarization matrix. $S$ is defined by $S=P\varepsilon^{-1}$ (matrix multiplication), where $P_{ij}$ and $\varepsilon_{ij}$ are the polarization and dielectric matrices in a tight binding representation, that we calculate following Ref.\cite{delerue97}. Eq. (\ref{eq:tb}) contains LFEs, and has been obtained within the RPA+LF approximation. When $S$ is substituted by $P$ (that is, when the polarization is unscreened), Eq. (\ref{eq:tb}) gives the standard independent particle RPA expression.

\begin{figure}
	\centering
	\includegraphics[width=0.45\textwidth]{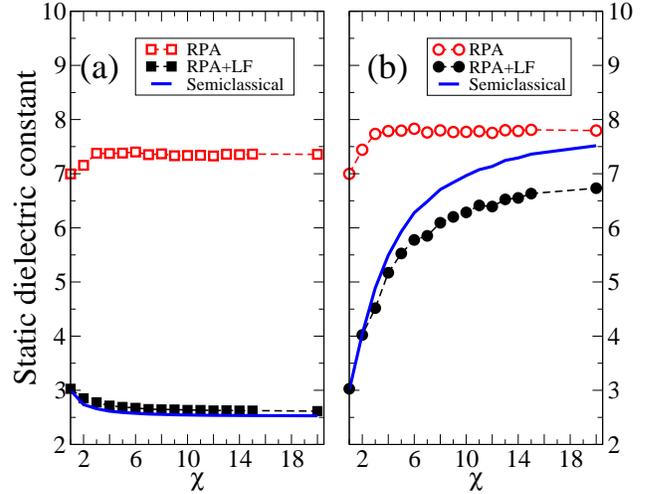}
	\caption{Static dielectric constant of Si ellipsoids with $a=0.5$ nm. Comparison between the RPA results (red-open symbols), the RPA+LF results (black-filled symbols), and the semiclassical model (blue-solid line). The perpendicular and parallel components (with respect to the ellipsoid axis) are shown in panels (a) and (b), respectively.}
	\label{fig:fig1}
\end{figure}

\section{Results} \label{sec:results}

We have studied Si ellipsoids having a rotational symmetry axis along the [001] crystallographic bulk Si direction. All the structures have been constructed with a Si atom in the center and passivated with hydrogen in order to avoid intra-gap states localized on the surface due to unsaturated dangling bonds. We define two semi-axes: $a$, lying in the plane orthogonal to the ellipsoid symmetry axis (the $z$-axis), and $c$, directed along the symmetry axis. The aspect ratio $\chi = c/a$ defines the geometric anisotropy of the ellipsoid. The spherical case corresponds to $\chi=1$, the quantum wire limit is obtained in the limit $\chi \to \infty$. All the details about the ellipsoids can be found in Refs.\cite{trani05,trani04}. Starting from the sphere, we study both the static dielectric constant and the absorption spectra of the ellipsoids, upon changing the aspect ratio. We keep $a$ fixed ($a = 0.5$ nm) and consider more and more elongated structures, up to reaching the quantum wire limit. We already performed RPA calculations on Si ellipsoids, taking into account QCEs\cite{trani05}. In the following we show the influence of LFEs, which are included using the RPA+LF method.

\begin{figure}
	\centering
	\includegraphics[width=0.45\textwidth]{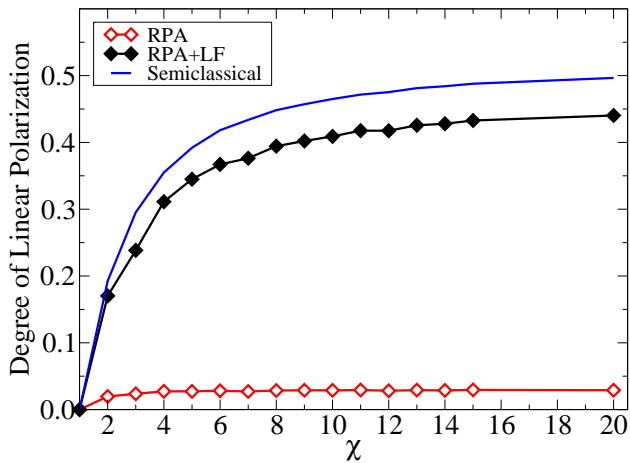}
	\caption{Degree of linear polarization for the static dielectric constant of Si ellipsoids with $a=0.5$ nm. Comparison between the RPA results (red-open symbols), the RPA+LF results (black-filled symbols) and the semiclassical model (blue-solid line).}
	\label{fig:fig2}
\end{figure}

In Fig. \ref{fig:fig1} the static dielectric constant is shown. The RPA and RPA+LF results are illustrated, together with the semiclassical dielectric model previously described. In the semiclassical model, QCEs are taken into account using the RPA result as the independent particle (Penn-like) dielectric constant, while SPEs are included using classical electrostatics.
For the perpendicular component (panel a), RPA+LF perfectly agrees with the semiclassical model, confirming that LFEs are due to the SPEs. The dielectric constant with SPEs included is much smaller than the RPA result, due to a strong depolarization effect. In both RPA and RPA+LF the curves are quite independent of the aspect ratio $\chi$, and they rapidly stabilize to their quantum wire limit.

For the parallel component a very different result emerges (panel b). Starting from the spherical case, the RPA+LF curve increases with the aspect ratio, tending to the RPA quantum wire limit for very elongated ellipsoids. Once again, the RPA+LF curve qualitatively agrees with the semiclassical model. We retrieve here the result, already known from the electrostatics, that for a quantum wire the parallel component of the electric field does not feel any depolarization effect. Therefore, starting from the sphere (isotropic case), where the depolarization effect is large, and increasing the aspect ratio, SPEs gradually decrease in the case of the parallel component, while they remain almost constant for the perpendicular case.

\begin{figure}
 \centering
   \includegraphics[height=0.45\textheight]{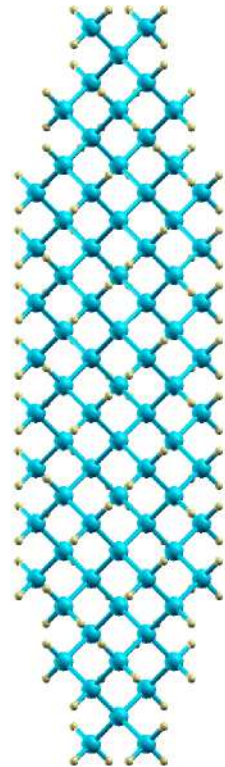}
 \caption{Example of a [001] ellipsoidal Si-nc. The $z$-axis is parallel to the ellipsoid symmetry axis. The in-plane semiaxis is $0.5$ nm, the aspect ratio is $4$. The big (blue) and the small (yellow) spheres respectively label Si and H atoms. In this case, the structure is constituted by $111$ Si and $124$ H atoms.}
 \label{fig:fig3}
\end{figure}

In Fig. \ref{fig:fig2} the degree of linear polarization for the static dielectric constant is shown. It is defined as the ratio between the difference and the sum of the parallel with respect to the perpendicular component, and gives an estimation of the dielectric anisotropy of the ellipsoids. A very interesting result coming out from Fig. \ref{fig:fig2} is that while QCEs give a very small anisotropy, less that $2\%$, the inclusion of LFEs dramatically enhances the anisotropy, that grows over $30\%$ for $\chi>4$. 
In Fig. \ref{fig:fig2} a discrepancy with the semiclassical case emerges. It is due to the fact that the dielectric model idealizes an ellipsoid as a perfectly regular and homogeneous structure. This is a fair approximation for quite regular structures, like spheres or slightly elongated ellipsoids. But, for huge aspect ratios and thin ellipsoids, the local atomistic nature gives non homogeneous features that are beyond the model.
In Fig. \ref{fig:fig3} an ellipsoid with aspect ratio $\chi=4$ is shown. We see that, upon increasing the aspect ratio the ellipsoid more closely resembles a cylinder with a local cap around the terminations. This could explain the underestimation of the depolarization effects found from the semiclassical model. 

\begin{figure}
 \centering
  \includegraphics[width=0.45\textwidth]{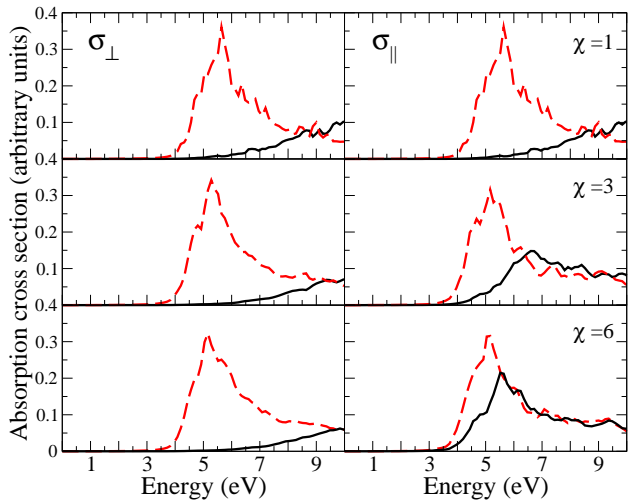}
 \caption{Absorption cross section for ellipsoidal Si-nc, upon increasing the aspect ratio $\chi$, at fixed in-plane semiaxis $a=0.5$ nm. The perpendicular (parallel) components with respect to the ellipsoid symmetry axis have been drawn on the left (right) side. The RPA and RPA+LF calculations are respectively shown with red (dashed) and black (solid) lines.}
 \label{fig:fig4}
\end{figure}

In Fig. \ref{fig:fig4} the absorption cross section is shown. The results already discussed for the static case well apply to the absorption cross section, too. The RPA+LF calculation shows that in the spherical case SPEs causes a strong annihilation of the absorption in the energy range from 3 to 7 eV. Upon increasing the aspect ratio, the perpendicular component is almost independent of the geometrical anisotropy. Instead, for the parallel component there is an increase of the absorption in the considered energy range. The main peak of the absorption spectra gradually increases and moves to lower energy, up to reaching the RPA limit for very elongated structures, in agreement with previous first principles calculations performed on quantum wires\cite{bruneval05}.

\section{Conclusion} \label{sec:conclusion}
In this paper the role of the local fields in the optical properties of silicon elllipsoids is discussed. It is shown that local field effects are responsible for a strong anisotropy in both the static dielectric constant and the absorption spectra. Local fields have been identified with the classical surface polarization effects, and the tight binding results have been compared to a dielectric model. While for spherical nanocrystals the agreement between the model and the atomistic results is very good, a discrepancy occurs for thin, elongated ellipsoids. This can be explained as a limit of the dielectric model, that does not take into account the atomistic features. From the absorption spectra important results emerge. While the perpendicular component is quite independent of the geometrical anisotropy, the parallel component shows a strong dependence. In particular, the energy of the main absorption peak changes and becomes smal ler and smaller upon increasing the aspect ratio, tending to the RPA peak energy for very elongated structures. Together with this, the intensity of the main peak increases with the aspect ratio. This signature should be experimentally observed, as it was in the case of carbon nanotubes\cite{li01}.

\ack
Financial support by COFIN-PRIN 2005 is acknowledged. Calculations have been partially performed at CINECA-``Progetti Supercalcolo 2006''.

 \end{document}